\documentclass[conference]{IEEEtran}
\IEEEoverridecommandlockouts
\ifCLASSINFOpdf
\else
   \usepackage{graphicx}
   \graphicspath{{../eps/}}
   \DeclareGraphicsExtensions{.eps}
\fi

\usepackage{amsmath}
\usepackage{amsfonts}
\usepackage{amssymb}
\usepackage{wrapfig}
\usepackage{psfrag}
\usepackage{epstopdf}
\usepackage{cite}
\usepackage{graphicx}
\usepackage{subfigure}
\usepackage{threeparttable}
\usepackage{cases}
\usepackage{subeqnarray}
\usepackage{color}
\usepackage{underscore}
\usepackage[algo2e,ruled,vlined]{algorithm2e}
\usepackage{algorithmic}

\usepackage{textcomp}
\usepackage{xcolor}
\def\BibTeX{{\rm B\kern-.05em{\sc i\kern-.025em b}\kern-.08em
    T\kern-.1667em\lower.7ex\hbox{E}\kern-.125emX}}
\begin{document}
\title{Enabling Full Mutualism for Symbiotic Radio with\\ Massive Backscatter Devices}

\author{{Jingran~Xu{*},~Zhuoyin~Dai{*},~and~Yong~Zeng{*$\dagger $}}\\
{{*}National~Mobile~Communications~Research~Laboratory,~Southeast~University,~Nanjing~210096,~China}\\
{{$\dagger $}Purple~Mountain~Laboratories,~Nanjing~211111,~China}\\
Email:~{$\text{ }\!\!\{\!\!\text{ }$jingran_xu,~zhuoyin_dai,~yong_zeng$\text{ }\!\!\}\!\!\text{ }$}@seu.edu.cn
}

\maketitle

 \vspace{-0.5cm}
\begin{abstract}
 Symbiotic radio is a promising technology to achieve spectrum- and energy-efficient wireless communications, where the secondary backscatter device (BD) leverages not only the spectrum but also the power of the primary signals for its own information transmission.  In return, the primary communication link can be enhanced by the additional multipaths created by the BD. This is known as the \emph{mutualism} relationship of symbiotic radio.
However, as the backscattering link is much weaker than the direct link due to double attenuations, the improvement of the primary link brought by one single BD is extremely limited. To address this issue and enable full mutualism of symbiotic radio, in this paper, we study symbiotic radio with massive number of BDs. For symbiotic radio multiple access channel (MAC) with successive interference cancellation (SIC), we first derive the achievable rate of both the primary and secondary communications, based on which a receive beamforming optimization problem is formulated and solved. Furthermore, considering the asymptotic regime of massive number of BDs, closed-form expressions are derived for the primary and the secondary communication rates, both of which are shown to be increasing functions of the number of BDs. This thus demonstrates that the mutualism relationship of symbiotic radio can be fully exploited with massive BD access.
\end{abstract}


% \fi
\IEEEpeerreviewmaketitle
 \vspace{-0.3cm}
\section{Introduction}
%As one of the promising technologies of the next generation Internet of things(IoT), symbiotic radio can save spectrum resources and reduce energy consumption effectively~\cite{IEEEhowto:1}, in which backscatter devices(BD) transmit their information over the incident primary signal via backscatter modulation instead of using active radio-frequency (RF) components~\cite{IEEEhowto:2},~\cite{IEEEhowto:3}.\\
Symbiotic radio has been recently proposed as a promising technology to achieve both spectrum- and energy-efficient wireless communications \cite{IEEEhowto:1,IEEEhowto:2,IEEEhowto:3}. For typical symbiotic radio systems, the secondary user utilizes the passive backscattering technology to transmit its own information. As such, different from the extensively studied cognitive radio systems \cite{IEEEhowto:4,IEEEhowto:5,IEEEhowto:6}, the secondary backscatter device (BD) in symbiotic radio systems leverages not only the spectrum but also the power of the primary signals for its own information transmission. Depending on the relations of the symbol durations of the primary and secondary signals, symbiotic radio can be classified into two categories \cite{IEEEhowto:7}, namely \emph{parasite symbiotic radio} (PSR) and \emph{commensal symbiotic radio} (CSR). In PSR, the secondary and primary signals have equal symbol durations, and the information transmission of BD introduces interferences to the primary transmission. By contrast, for CSR, the symbol duration of BD signals is much longer than that of the primary signals. As a result, rather than causing interference, the backscattering of the BD actually creates additional multipaths that can be exploited to enhance the primary communication link. This is known as the \emph{mutualism} relationship of symbiotic radio \cite{IEEEhowto:7}, which makes it especially appealing for spectrum- and energy-efficient Internet of Things (IoT) networks.

Significant research efforts have been recently devoted to the theoretical analysis and practical design of symbiotic radio systems. For instance, the performance analysis in terms of achievable rate \cite{IEEEhowto:8,IEEEhowto:9} and outage probability \cite{IEEEhowto:10} are given in different systems. The practical receiver design of symbiotic radio systems is considered in \cite{IEEEhowto:11,IEEEhowto:12,IEEEhowto:13}. For multi-antenna symbiotic radio systems, the beamforming optimization has been extensively studied to maximize various performance metrics, e.g., energy efficiency \cite{IEEEhowto:8}, sum capacity of the primary and secondary communications \cite{IEEEhowto:14}, or fairness of the secondary users \cite{IEEEhowto:15}.

Despite of the promising benefits, symbiotic radio also faces new critical challenges. In particular, as the backscattered signal suffers from double power attenuation, the strength of the backscattering link is usually much weaker than that of the direct link between the primary transmitter (PT) and primary receiver. This not only limits the communication performance of the secondary system itself, but also compromises its promised performance enhancement to the primary communication links. There are some preliminary efforts to address such issues, e.g., via active load [16]--[17] or employing large reconfigurable intelligent surface (RIS) [18]--[19] at the secondary devices. Though effective for enhancing the secondary links, such methods may drastically increase the power, cost, complexity, and size of the secondary devices, which may undermine the initial motivation of symbiotic radio.

In this paper, we propose an alternative method to significantly enhance the secondary backscattering links and enable the full mutualism of symbiotic radio systems, by exploiting the potential gain brought by multiple BDs. This is motivated by the 6G visions to support ultra-massive connectivity, say 10 million devices per square kilometers \cite{IEEEhowto:20}, most of which are expected to be IoT devices. This thus provides abundant multi-user diversities for symbiotic radio to not only achieve high performance for secondary communications, but also offer significant enhancement to primary communications. This thus motivates our current work to study symbiotic radio with massive BDs.
To that end, we first derive the achievable rate expression of the primary communication, as well as the sum rate expression of all BDs, by noting that it corresponds to a multiple access channel (MAC), where minimum mean square error estimation (MMSE) with successive interference cancellation (SIC) is optimal \cite{IEEEhowto:21}.
Furthermore, an optimization problem is formulated to maximize the primary communication rate via receive beamforming, by taking into account both the direct primary link and the additional multipaths created by the BDs. The problem is non-convex and two solutions that are applicable to different scenarios and with different complexity are proposed, namely \emph{correlation matrix based solution} and \emph{closed-form based semi-definite relaxation (SDR) solution}.
To gain useful insights, we further study the asymptotic regime of massive number of BDs and derive closed-form expressions for the primary and the secondary communication rates, both of which are shown to be increasing functions of the number of BDs. This thus demonstrates that the mutualism relationship of symbiotic radio can be fully enabled with massive BD access.
\section{System Model}
\begin{figure}[!t]
  \centering
  \centerline{\includegraphics[width=3.5in,height=2.0in]{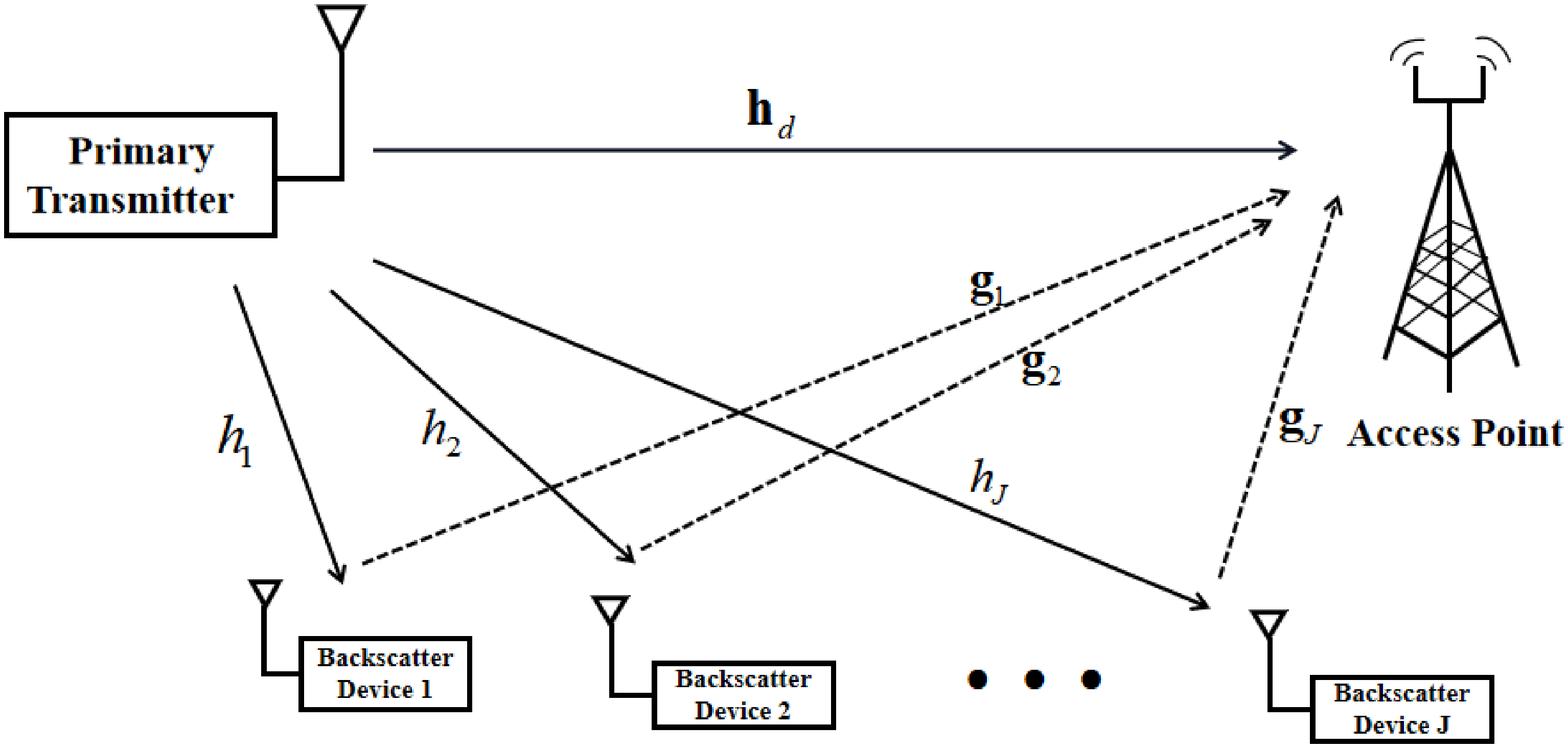}}
  \caption{Symbiotic radio with massive backscatter devices.}
  \label{system model}
   \vspace{-0.5cm}
  \end{figure}
As shown in Fig. 1, we consider a multi-BD symbiotic radio system, which consists of one PT, one access point (AP) and $J$ BDs. We assume that the AP has $M$ antennas, and the PT and each BD are equipped with one antenna. The AP wishes to decode not only the primary information from the PT, but also the secondary information from the $J$ BDs, which modulate their information via backscattering the incident primary signal by intelligently varying their reflection coefficients. As such, the BDs share not only the same spectrum but also the power with the PT. In return, their scattered signals may create additional channel paths to enhance the primary communication link, as long as their symbol rate is much lower than that of the primary signal. This is known as \emph{mutualism} relationship \cite{IEEEhowto:7}.
We denote the single-input multiple-output (SIMO) direct-link channel from the PT to the AP as ${{\mathbf{h}}_{d}}\in {{\mathbb{C}}^{M\times 1}}$. Further denote by  ${{h}_{j}}\in \mathbb{C}$ the channel coefficient from the PT to BD $j$, and $\mathbf g_j\in \mathbb{C}^{M\times 1}$ the SIMO channel from BD $j$ to the AP. Then the cascaded backscattering link coefficient from the PT to AP via BD $j$ is $h_j\mathbf g_j$.

We focus on the CSR setup \cite{IEEEhowto:7}, where the symbol rate of the PT is $K\gg 1$ times of that of the BDs. In other words, over each BD symbol duration, $K$ PT symbols can be transmitted. Let $c_j(n)$ denote the information-bearing symbol of BD $j$, and $s(k,n)$ denote the information-bearing symbols of the PT, where $k=1,....,K$. We assume that $s(k,n)$ follows the independent and identically distributed (i.i.d.) circularly symmetric complex gaussian (CSCG) distribution, i.e., $s(k,n)\sim \mathcal{C}\mathcal{N}(0,1)$. Furthermore, let $p$ denote the transmit power by the PT, and $\alpha \in \left[ 0,1 \right]$ be the fraction of the power backscattered by each BD.
Then the signal received by the AP during the $n$th BD symbol duration is
\begin{equation}
\setlength\abovedisplayskip{1pt}
\setlength\belowdisplayskip{1pt}
 \begin{aligned}
\hspace{-3ex}\mathbf{y}(k,n)&=\sqrt{p}{{\mathbf{h}}_{d}}s(k,n)+\sum\limits_{j=1}^{J}{\sqrt{p}}\sqrt{\alpha }{{h}_{j}}{{\mathbf{g}}_{j}}s(k,n){{c}_{j}}(n)\\&+\mathbf{z}(k,n),k=1,...,K,
 \end{aligned}
\end{equation}
where $\mathbf{z}(k,n)\in \mathbb{C}^{M\times 1}$ is the i.i.d. CSCG noise with zero mean and power ${{\sigma }^{2}}$, i.e., $\mathbf{z}(k,n)\sim\mathcal{C}\mathcal{N}\left( \mathbf 0, \sigma^2{\mathbf{I}}_{M}\right)$.

% >>>>>>>>>>>>>SECTIONS II -  here >>>>>>>>>>>>
\vspace{-0.1cm}
\section{Achievable Rate Analysis}
To decode the primary signal $s(k,n)$ from the $M$-dimensional signal $\mathbf y(k,n)$ in (1), the AP applies a receive beamforming vector $\mathbf{w}_{d}$, where $\left\| \mathbf{w}_{d} \right\|=1$, which gives
\setlength\abovedisplayskip{2pt}
\begin{equation}
\setlength\abovedisplayskip{1pt}
\setlength\belowdisplayskip{1pt}
 \begin{aligned}
y(k,n)& =\sqrt{p}\mathbf{w}_{d}^{\text{H}}{{\mathbf{h}}_{d}}s(k,n)\\&+\sum\limits_{j=1}^{J}{\sqrt{p}}\sqrt{\alpha }\mathbf{w}_{d}^{\text{H}}{{h}_{j}}{{\mathbf{g}}_{j}}s(k,n){{c}_{j}}(n)+\mathbf{w}_{d}^{\text{H}}\mathbf{z}(k,n).\\
 \end{aligned}
\end{equation}
\setlength\belowdisplayskip{2pt}Since the BD symbols $c_j(n)$ remain unchanged for each block of $K$ PT symbols, the second term in (2) constitutes the additional multi-path channels for the primary signal. As a result, the equivalent SIMO channel for decoding $s(k,n)$ is dependent on the BD symbols $\mathbf c(n)=[c_1(n),c_2(n),...,c_J(n)]^{\text{T}}$, which is denoted as ${{\mathbf{h}}_{eq}}(\mathbf{c}(n))={{\mathbf{h}}_{d}}+\sum\limits_{j=1}^{J}{\sqrt{\alpha }}{{h}_{j}}{{\mathbf{g}}_{j}}{{c}_{j}}(n)$.
Given $\mathbf{c}(n)$, the SNR for the primary signal is
 \begin{equation}
 \setlength\abovedisplayskip{0.1cm}
\setlength\belowdisplayskip{0.1cm}
{{{r}_{s}}(\mathbf{c}(n))=\frac{p{{\left| \mathbf{w}_{d}^{\text{H}}{{\mathbf{h}}_{eq}} (\mathbf{c}(n)) \right|}^{2}}}{{{\sigma }^{2}}}}.
 \end{equation}
For sufficiently large $K$, the average primary rate is~\cite{IEEEhowto:7}
 \begin{equation}
 \setlength\abovedisplayskip{0.1cm}
\setlength\belowdisplayskip{0.1cm}
{{{R}_{s}}={{\mathbb{E}}_{\mathbf{c}(n)}}\left[ {{\log }_{2}}(1+{{r}_{s}}(\mathbf{c}(n))) \right]}.
 \end{equation}
On the other hand, to decode the BD symbols $c_j(n)$ for each $n$, by concatenating $\mathbf{y}(k,n)$ in (1) for all $k=1,2,\cdots ,K$, we have
$\mathbf{Y}(n)=\left[ \mathbf{y}(1,n),\mathbf{y}(2,n),\cdots ,\mathbf{y}(K,n) \right]\in \mathbb{C}^{M\times K}$.
Similarly, let $\mathbf{s}(n)={{\left[ s(1,n),s(2,n),\cdots s(K,n) \right]}^{\text{T}}}\in \mathbb{C}^{K \times 1}$ and $\mathbf{Z}(n)=\left[ \mathbf{z}(1,n),\mathbf{z}(2,n),\cdots ,\mathbf{z}(K,n) \right]\in  \mathbb{C}^{M\times K}$.
Then (1) can be compactly written as
\begin{small}
\begin{equation}
\setlength\abovedisplayskip{2pt}
\setlength\belowdisplayskip{2pt}
{\hspace{-3ex}\mathbf{Y}(n)=\sqrt{p}{{\mathbf{h}}_{d}}{\mathbf{s}}^{\text{T}}{(n)}+\sum\limits_{j=1}^{J}{\sqrt{p}}\sqrt{\alpha }{{h}_{j}}{{\mathbf{g}}_{j}}{\mathbf{s}}^{\text{T}}{(n)}{{c}_{j}}(n)+\mathbf{Z}(n)}.
\end{equation}
\end{small}After decoding $s(k,n)$, the primary signal component can be subtracted from (5) before decoding the BD signals, which results
\begin{equation}
\setlength\abovedisplayskip{1pt}
\setlength\belowdisplayskip{1pt}
{\mathbf{\hat{Y}}(n)=\sum\limits_{j=1}^{J}{\sqrt{p}}\sqrt{\alpha }{{h}_{j}}{{\mathbf{g}}_{j}}{\mathbf{s}}^{\text{T}}{(n)}{{c}_{j}}(n)+\mathbf{Z}(n)}.
 \end{equation}
Furthermore, the optimal temporal-domain matched filtering can be applied, by right multiplying $\mathbf{\hat{Y}}(n)$ in (6) by $\mathbf{v}(n)=\mathbf{s}{{(n)}^{*}}/{\left\| \mathbf{s}(n) \right\|}.$
The resulting signal is
\begin{equation}
\setlength\abovedisplayskip{1pt}
\setlength\belowdisplayskip{1pt}
{\mathbf{\hat{y}}(n)=\sum\limits_{j=1}^{J}{\sqrt{p}}\sqrt{\alpha }{{h}_{j}}{{\mathbf{g}}_{j}}\left\| \mathbf{s}(n) \right\|{{c}_{j}}(n)+\mathbf{Z}(n)\mathbf{v}(n)}.
\end{equation}
For sufficiently large $K$, due to the law of large numbers, ${{\left\| \mathbf{s}(n) \right\|}^{2}}$ approaches $K$.
Therefore, $\mathbf{\hat{y}}(n)$ in (7) approaches to
\begin{small}
\begin{equation}
\setlength\abovedisplayskip{1pt}
\setlength\belowdisplayskip{1pt}
{\mathbf{\hat{y}}(n)=\sqrt{Kp\alpha}\sum\limits_{j=1}^{J}{{{h}_{j}}{{\mathbf{g}}_{j}}{{c}_{j}}(n)}+\hat{\mathbf z}(n)},
\end{equation}
\end{small}where $\hat{\mathbf z}(n)=\mathbf{Z}(n)\mathbf{v}(n)\sim \mathcal{CN}(\mathbf 0, \sigma^2\mathbf I_M)$.
Note that (8) is essentially a SIMO MAC, where MMSE-SIC receiver is known to be capacity-achieving~\cite{IEEEhowto:21}. Specifically, the $J$ BD users are ordered according to their channel strength ${{\left\| {{h}_{j}}{{\mathbf{g}}_{j}} \right\|}^{2}}$, based on which the SIC decoding order is determined.
Without loss of generality, assume that ${{\left\| {{h}_{1}}{{\mathbf{g}}_{1}} \right\|}^{2}}\ge {{\left\| {{h}_{2}}{{\mathbf{g}}_{2}} \right\|}^{2}}\ge \cdots \ge {{\left\| {{h}_{J}}{{\mathbf{g}}_{J}} \right\|}^{2}}$, then the SIC decoding order is $1,2,\cdots ,J$.
Let's focus on BD $j$, where the signals for BDs $1,...,j-1$ have already been decoded and perfectly removed, and those for BDs $j+1,...,J$ are treated as noise. Denote the spatial-domain beamforming vector for BD $j$ as ${{\mathbf{w}}_{j}}\in {{\mathbb{C}}^{M\times 1}}$. Then the resulting signal can be written as
%\begin{small}
\begin{equation}
\setlength\abovedisplayskip{2pt}
\setlength\belowdisplayskip{2pt}
 \begin{aligned}
{{y}_{j}}(n)&=\sqrt{Kp\alpha}{{h}_{j}}\mathbf{w}_{j}^{\text{H}}{{\mathbf{g}}_{j}}{{c}_{j}}(n)\\&+\sqrt{Kp\alpha}\mathbf{w}_{j}^{\text{H}}\sum\limits_{i=j+1}^{J}{{{h}_{i}}{{\mathbf{g}}_{i}}{{c}_{i}}(n)}+\mathbf{w}_{j}^{\text{H}}\hat{\mathbf z}(n).
 \end{aligned}
\end{equation}
%\end{small}
The linear MMSE beamforming that maximizes the SINR is
\begin{equation}
\setlength\abovedisplayskip{2pt}
\setlength\belowdisplayskip{2pt}
{{{\mathbf{w}}_{j}}={{\Big( Kp\alpha \sum\limits_{i=j+1}^{J}{{{\left| {{h}_{i}} \right|}^{2}}{{\mathbf{g}}_{i}}}\mathbf{g}_{i}^{\text{H}}+{{\sigma }^{2}}{{\mathbf{I}}_{M}} \Big)}^{-1}}\sqrt{Kp\alpha }{{h}_{j}}{{\mathbf{g}}_{j}}}.
\end{equation}
The corresponding maximum SINR is
\begin{equation}
\setlength\abovedisplayskip{2pt}
\setlength\belowdisplayskip{2pt}
{\hspace{-3ex}{{\gamma }_{{{c}_{j}}}}=Kp\alpha {{\left| {{h}_{j}} \right|}^{2}}\mathbf{g}_{j}^{\text{H}}{{\Big( Kp\alpha \sum\limits_{i=j+1}^{J}{{{\left| {{h}_{i}} \right|}^{2}}{{\mathbf{g}}_{i}}}\mathbf{g}_{i}^{\text{H}}+{{\sigma }^{2}}{{\mathbf{I}}_{M}}\Big)}^{-1}}{{\mathbf{g}}_{j}}}.
\end{equation}
As a result, the sum rate of the $J$ BDs can be written as
\begin{equation}
\setlength\abovedisplayskip{2pt}
\setlength\belowdisplayskip{2pt}
{{R}_{BD}={\frac{1}{K}}\sum\limits_{j=1}^{J}{\log }_{2} \left( 1+{{\gamma }_{{{c}_{j}}}} \right)}.
\end{equation}
It can be shown that the sum rate in (12) can also be expressed as [21, chapter 8]:
%\begin{small}
\begin{equation}
\setlength\abovedisplayskip{2pt}
\setlength\belowdisplayskip{2pt}
{{R}_{BD}=\frac{1}{K}{\log }_{2} \det \left( \mathbf{I}_M+\frac{Kp\alpha }{{{\sigma }^{2}}}\sum\limits_{j=1}^{J}{{{\left| {{h}_{j}} \right|}^{2}}{{\mathbf{g}}_{j}}\mathbf{g}_{j}^{\text{H}}} \right)}.
\end{equation}
%\end{small}

To show how the sum rate (13) is affected by the number of BDs $J$, we consider the asymptoic performance for massive BDs, i.e., as $J$ goes sufficiently large. To this end, we assume that the BD channels are i.i.d. distributed, with $ \mathbb{E}[ {{\left| {{h}_{j}} \right|}^{2}}]={{\beta }_{h}}$ and $\mathbb{E}\left[ {{\mathbf{g}}_{j}}\mathbf{g}_{j}^{\text{H}} \right]={{\beta }_{g}}{{\mathbf{I}}_{M}}$, $\forall j=1,...,J$, where $\beta_h$ and $\beta_g$ are the average channel gains. We then have the following Lemma:

\emph{Lemma 1:} For symbiotic radio with massive BDs, i.e., $J\gg  1$, $R_{BD}\to \frac{M}{K}{\log }_{2} \left( 1+\frac{JKp\alpha {{\beta }_{h}}{{\beta }_{g}}}{{{\sigma }^{2}}} \right)$.

\begin{IEEEproof}
Due to the law of large numbers, for $J\gg 1$, we have:
$\sum\limits_{j=1}^{J}{{{\left| {{h}_{j}} \right|}^{2}}}{{\mathbf{g}}_{j}}\mathbf{g}_{j}^{\text{H}}\to J{{\beta }_{h}}{{\beta }_{g}}{{\mathbf{I}}_{M}}$.
It then follows from (13) that
%\begin{small}
\begin{equation}
\setlength\abovedisplayskip{2pt}
\setlength\belowdisplayskip{2pt}
\begin{split}
   R_{BD}& \to \frac{1}{K}{\log }_{2} \det \left( \mathbf{I}_M+\frac{JKp\alpha {{\beta }_{h}}{{\beta }_{g}}{{\mathbf{I}}_{M}}}{{{\sigma }^{2}}} \right) \\
 & \to \frac{M}{K}{\log }_{2} \left( 1+\frac{JKp\alpha {{\beta }_{h}}{{\beta }_{g}}}{{{\sigma }^{2}}} \right). \\
\end{split}
\end{equation}
%\end{small}
\end{IEEEproof}
Lemma 1 shows that for symbiotic radio with massive BDs, the sum rate of the BDs increases monotonically with the number of BDs $J$, thanks to the multi-user diversity gains.

%%>>>>>>>>>>>>>> Section III  >>>>>>>>>>>>>>>>>>>>
\section{Beamforming Optimization}
In this section, the receive beamforming $\mathbf w_d$ is optimized for the primary communications, based on the rate expression (4). The following problem can be formulated:
\begin{equation}
\setlength\abovedisplayskip{0.2cm}
\setlength\belowdisplayskip{0.2cm}
{\hspace{-3ex}\left( \text{P1} \right)\underset{{{\left\| {{\mathbf{w}}_{d}} \right\|}}=1}{\mathop{\max }}\,\text{ }{{R}_{s}}={{\mathbb{E}}_{\mathbf{c}(n)}}\Big[ {{\log }_{2}}\Big( 1+\frac{p{{\left| \mathbf{w}_{d}^{\text{H}}{{\mathbf{h}}_{eq}}\left( \mathbf{c}(n) \right) \right|}^{2}}}{{{\sigma }^{2}}} \Big) \Big]}.
\end{equation}
Note that $\mathbf{w}_{d}$ in (P1) needs to be optimized to maximize the expected primary communication rate, with the expectation taken with respect to the random BD symbols $\mathbf c(n)$. In the trivial scenario that $\mathbf c(n)$ is deterministic, denoted by $\mathbf {\bar c}$, it is obvious that the optimal receive beamforming is the matched filter to the equivalent channel ${\mathbf{h}}_{eq}(\mathbf{\bar c})$, i.e, ${{\mathbf{w}}_{d}}=\frac{{{\mathbf{h}}_{eq}}(\mathbf{\bar{c}})}{\left\| {{\mathbf{h}}_{eq}}(\mathbf{\bar{c}}) \right\|}$. However, for the general case where $\mathbf c(n)$ is random, the solution to (P1) is non-trivial due to its non-convexity nature. In the following, we propose two solutions to (P1), termed \emph{correlation matrix based solution} and \emph{closed-form based SDR solution}, which are applicable to different scenarios and with different complexity.
\subsection{Correlation Matrix Based Solution}

%As the sample-average SDR method involves the sample average approximation and Gaussian randomization, there is no guarantee that the resulting solution is globally optimal. Besides, it involves solving the SDP problem, which incurs high complexity as $M$ goes large. To further reduce the computation complexity,
With correlation matrix based solution, $R_s$ in (15) is approximated by its upper bound obtained by using Jensen's inequality to the concave logarithmic function, i.e.,
\begin{equation}
\setlength\abovedisplayskip{2pt}
\setlength\belowdisplayskip{2pt}
\begin{split}
  {R_s\leq{R}_{{{s}_{UB}}}}& \triangleq{{\log }_{2}}\Big( 1+\frac{p{{\mathbb{E}}_{\mathbf{c}(n)}}\left[ {{\left| \mathbf{w}_{d}^{\text{H}}{{\mathbf{h}}_{eq}}(\mathbf{c}(n)) \right|}^{2}} \right]}{{{\sigma }^{2}}} \Big) \\
 & ={{\log }_{2}}\left( 1+\frac{p\mathbf{w}_{d}^{\text{H}}\mathbf{M}{{\mathbf{w}}_{d}}}{{{\sigma }^{2}}} \right), \\
\end{split}
\end{equation}where $\mathbf{M}$ is the correlation matrix of ${{\mathbf{h}}_{eq}}(\mathbf{c}(n))$ given by
\begin{small}
\begin{equation}
\setlength\abovedisplayskip{2pt}
\setlength\belowdisplayskip{2pt}
\hspace{-3ex}\begin{split}
  & \mathbf{M} ={{\mathbb{E}}_{\mathbf{c}(n)}}\left[ {{\mathbf{h}}_{eq}}(\mathbf{c}(n))\mathbf{h}_{eq}^{\text{H}}(\mathbf{c}(n)) \right] \\
 & ={{\mathbb{E}}_{\mathbf{c}(n)}}\left[ \left( {{\mathbf{h}}_{d}}+\sum\limits_{j=1}^{J}{\sqrt{\alpha }}{{h}_{j}}{{\mathbf{g}}_{j}}{{c}_{j}}(n) \right){{\left( {{\mathbf{h}}_{d}}+\sum\limits_{j=1}^{J}{\sqrt{\alpha }}{{h}_{j}}{{\mathbf{g}}_{j}}{{c}_{j}}(n) \right)}^{\text{H}}} \right] \\
 & ={{\mathbf{h}}_{d}}\mathbf{h}_{d}^{\text{H}}+\alpha {{\mathbb{E}}_{\mathbf{c}(n)}}\left[ \sum\limits_{i=1}^{J}{{\sum\limits_{j=1}^{J}{{{c}_{i}}(n){{\left( {{c}_{j}}(n) \right)}^{*}}{{h}_{i}}{{\mathbf{g}}_{i}}{\left( {{h}_{j}}{{\mathbf{g}}_{j}} \right)}}^{\text{H}}}} \right] \\
 & ={{\mathbf{h}}_{d}}\mathbf{h}_{d}^{\text{H}}+\alpha \sum\limits_{j=1}^{J}{{{\left| {{h}_{j}} \right|}^{2}}{{\mathbf{g}}_{j}}\mathbf{g}_{j}^{\text{H}}},
\end{split}
\end{equation}
\end{small}where the last equality follows since ${{c}_{j}}(n)$ are i.i.d. random symbols for different BDs $j$.
Therefore, by replacing $R_s$ in (15) with (16) and ignoring constant terms, we have
\begin{equation}
\setlength\abovedisplayskip{0.2cm}
\setlength\belowdisplayskip{0.2cm}
(\text{P}2)\underset{{{\left\| {{\mathbf{w}}_{d}} \right\|}}=1}{\mathop{\max }}\,\text{ }\mathbf{w}_{d}^{\text{H}}\mathbf{M}{{\mathbf{w}}_{d}}.
\end{equation}
Apparently, the optimal solution to (P2) is given by the dominant eigenvector of the positive semidefinite matrix of $\mathbf{M}$.
\subsection{Closed-form Based SDR Solution}
The correlation matrix based solution is applicable to any distribution of the BD symbols $\mathbf c(n)$. In this subsection, we propose an alternative solution for the special case when ${{c}_{j}}(n)$ are i.i.d. CSCG distributed, i.e., ${{c}_{j}}(n)\sim\mathcal{C}\mathcal{N}(0,1)$, for which a semi-closed form expression of the expected primary rate (4) can be obtained. Specifically, with CSCG BD symbols, ${{r}_{s}}(\mathbf{c}(n))$ in (3) follows a noncentral chi-square distribution ${{\chi }^{2}}$ with the freedom of 2 \cite{IEEEhowto:7}, the non-centrality parameter $\lambda =\text{Re}{{\left\{ \frac{\sqrt{p}}{\sigma }\mathbf{w}_{d}^{\text{H}}{{\mathbf{h}}_{d}} \right\}}^{2}}+\operatorname{Im}{{\left\{ \frac{\sqrt{p}}{\sigma }\mathbf{w}_{d}^{\text{H}}{{\mathbf{h}}_{d}} \right\}}^{2}}=\frac{p{{\left| \mathbf{w}_{d}^{\text{H}}{{\mathbf{h}}_{d}} \right|}^{2}}}{{{\sigma }^{2}}}$, and the Gaussian variance parameter $\Sigma =\sum\limits_{j=1}^{J}{\frac{p\alpha {{\left| {{h}_{j}} \right|}^{2}}{{\left| \mathbf{w}_{d}^{\text{H}}{{\mathbf{g}}_{j}} \right|}^{2}}}{2{{\sigma }^{2}}}}$.
As a result, the probability density function (PDF) of $r_s(\mathbf c(n))$ in (3) is given by
 \begin{equation}
 \setlength\abovedisplayskip{0.2cm}
\setlength\belowdisplayskip{0.2cm}
{f(x)=\frac{1}{2\Sigma }{{e}^{\left( -\frac{x+\lambda }{2\Sigma } \right)}}{{I}_{0}}(\frac{\sqrt{x\lambda}}{\Sigma })},
 \end{equation}
where ${{I}_{0}}(\cdot )$ is a modified Bessel function of the first kind.
When the SNR ${{r}_{s}}(\mathbf{c}(n))$ is sufficient large, following similar derivation as~\cite{IEEEhowto:7}, the expected rate in (4) can be written as
\begin{equation}
\setlength\abovedisplayskip{0.3cm}
\setlength\belowdisplayskip{2pt}
\begin{split}
   {{R}_{s}}&={{\mathbb{E}}_{\mathbf{c}(n)}}\left[{{\log }_{2}}(1+{{r}_{s}}(\mathbf{c}(n))) \right]\\
  &\approx {{\mathbb{E}}_{\mathbf{c}(n)}}\left[ {{\log }_{2}}({{r}_{s}}(\mathbf{c}(n)))\right]\\
 & =({{\log }_{2}}e)\int_{0}^{\infty }{(\ln x)\frac{1}{2\Sigma }}{{e}^{\left( -\frac{x+\lambda }{2\Sigma } \right)}}{{I}_{0}}(\frac{\sqrt{x\lambda }}{\Sigma })\text{d}x \\
 & ={{\log }_{2}}\lambda -{\text{Ei}}(-\frac{\lambda }{2\Sigma }){{\log }_{2}}e,
 \vspace{-0.3cm}\label{eq:RS}
\end{split}
\end{equation}
where $\text{Ei}(x)\triangleq \int_{-\infty }^{x}{\frac{{{e}^{t}}}{t}}\text{d}t$ is the exponential integral. The expression \eqref{eq:RS} shows that for symbiotic radio system, the expected primary rate is given by a summation of the rate achievable by the direct link, i.e., $\log_2\lambda$, and an additional rate gain $\Delta {{R}_{s}}=-{\text{Ei}}\left( -\frac{\lambda }{2\Sigma } \right){{\log }_{2}}e>0$. Such a result was firstly revealed in~\cite{IEEEhowto:7} for symbiotic radio systems with one BD, and \eqref{eq:RS} shows that it is also applicable to multi-BD symbiotic radio systems, where the variance parameter $\Sigma$ is given by the aggregated contributions from all the $J$ BDs. It is not difficult to see that for any given direct link SNR $\lambda$, $R_s$ in \eqref{eq:RS} monotonically increases with $\Sigma$. Thus, with more BDs connected to the symbiotic radio system, the enhancement to the primary transmission becomes more significant.
%The expected value of the logarithm of a non-central chi-square random variable $v$ with an even number 2$m$ of degrees of freedom is given as~\cite{IEEEhowto:21}\\
% \begin{equation}
%{{\mathop{\rm E}\nolimits} [\ln v] = {q_m}({s^2})}
% \end{equation}
%  \begin{equation}
%   \begin{aligned}
% {{q}_{m}}\triangleq \ln x-{\text{Ei}}(-x)+\sum\limits_{j=1}^{m-1}{{{(-1)}^{j}}}[{{e}^{-x}}(j-1)!\\-\frac{(m-1)!}{j(m-1-j)!}]{{(\frac{1}{x})}^{j}},x>0
% \end{aligned}
% \end{equation}
%  \begin{equation}
%{{q_m}({s^2}) = \int_0^\infty  {\ln v}\cdot {(\frac{v}{{{s^2}}})^{\frac{{m - 1}}{2}}}{e^{ - v - {s^2}}}{{I}_{m-1}}(2s\sqrt v ){\mathop{\text{d}v}\nolimits}}
% \end{equation}
%Applying the linear transformation $v=\frac{x}{2\Sigma },{{s}^{2}}=\frac{\lambda }{2\Sigma }$ to (29), we have
%\begin{equation}
%\begin{split}
%  & {{R}_{s}}=({{\log }_{2}}e)\int_{0}^{\infty }{\ln v\cdot {{e}^{(-v-{{s}^{2}})}}}{{I}_{0}}(2s\sqrt{v})\text{d}v \\
% & \text{       }+({{\log }_{2}}(2\Sigma ))\int_{0}^{\infty }{{{e}^{(-v-{{s}^{2}})}}}{{I}_{0}}(2s\sqrt{v})\text{d}v \\
% & \text{   }={{\log }_{2}}e\cdot {{q}_{1}}({{s}^{2}})+{{\log }_{2}}(2\Sigma ) \\
% & \text{   }={{\log }_{2}}e\cdot {{q}_{1}}(\frac{\lambda }{2\Sigma })+{{\log }_{2}}(2\Sigma ) \\
% & \text{   }={{\log }_{2}}\lambda -{\text{Ei}}(-\frac{\lambda }{2\Sigma }){{\log }_{2}}e \\
%\end{split}
%\end{equation}

To gain further insight, we study the asymptotic performance of $R_s$ in \eqref{eq:RS} for massive BDs, i.e., $J\gg 1$. Similar to Lemma 1, when $h_j$ and ${{\mathbf{g}}_{j}}$ are i.i.d. channels with average channel gains $\beta_h$ and $\beta_g$ respectively, we have the following Lemma:

\emph{Lemma 2:} For symbiotic radio with massive BDs, i.e., $J\gg 1$, we have
\begin{equation}
\setlength\abovedisplayskip{0.2cm}
\setlength\belowdisplayskip{0.2cm}
{{{R}_{s}}\to {{\log }_{2}}\frac{p{{\left| \mathbf{w}_{d}^{\text{H}}{{\mathbf{h}}_{d}} \right|}^{2}}}{{{\sigma }^{2}}}-\text{Ei}\left( -\frac{{{\left| \mathbf{w}_{d}^{\text{H}}{{\mathbf{h}}_{d}} \right|}^{2}}}{J\alpha {{\beta }_{h}}{{\beta }_{g}}} \right){{\log }_{2}}e}.
\end{equation}
\begin{IEEEproof}
Due to the law of large numbers, we have:
%\begin{small}
\begin{equation}
\setlength\abovedisplayskip{0.2cm}
\setlength\belowdisplayskip{0.2cm}
\begin{split}
   \Sigma & =\frac{p\alpha }{2{{\sigma }^{2}}}\mathbf{w}_{d}^{\text{H}}\left( \sum\limits_{j=1}^{J}{{{\left| {{h}_{j}} \right|}^{2}}{{\mathbf{g}}_{j}}\mathbf{g}_{j}^{\text{H}}} \right){{\mathbf{w}}_{d}}\\
   & \to \frac{p\alpha }{2{{\sigma }^{2}}}\mathbf{w}_{d}^{\text{H}}J{{\beta }_{h}}{{\beta }_{g}}{{\mathbf{I}}_{M}}{{\mathbf{w}}_{d}} \\
 & =\frac{Jp\alpha {{\beta }_{h}}{{\beta }_{g}}}{2{{\sigma }^{2}}}. \\
\end{split}
\end{equation}
%\end{small}
By subsituting (22) into \eqref{eq:RS}, Lemma 2 then follows.
\end{IEEEproof}

Similar to \eqref{eq:RS}, it is not difficult to see that $R_s$ in (21) increases monotonically with the number of BDs $J$.

\emph{Lemma 3:} For symbiotic radio with massive BDs, i.e., $J\gg 1$, the optimal beamforming $\mathbf{w}_{d}$ that maximizes $R_s$ in (21) is $\mathbf{w}_{d}=\frac{{{\mathbf{h}}_{d}}}{\left\| {{\mathbf{h}}_{d}} \right\|}$, and the resulting primary rate is $R_s={{\log }_{2}}\frac{p{{\left\| {{\mathbf{h}}_{d}} \right\|}^{2}}}{{{\sigma }^{2}}}-\text{Ei}\left( -\frac{{{\left\| {{\mathbf{h}}_{d}} \right\|}^{2}}}{J\alpha {{\beta }_{h}}{{\beta }_{g}}} \right){{\log }_{2}}e$.
\begin{IEEEproof}
By taking the derivative of $R_s$ in (21) with respect to ${{\left| \mathbf{w}_{d}^{\text{H}}{{\mathbf{h}}_{d}} \right|}^{2}}$ as a whole, we have $\frac{\partial {{R}_{s}}}{\partial \left( {{\left| \mathbf{w}_{d}^{\text{H}}{{\mathbf{h}}_{d}} \right|}^{2}} \right)}=\frac{{{\log }_{2}}e}{{{\left| \mathbf{w}_{d}^{\text{H}}{{\mathbf{h}}_{d}} \right|}^{2}}}( 1-{{e}^{\frac{-{{\left| \mathbf{w}_{d}^{\text{H}}{{\mathbf{h}}_{d}} \right|}^{2}}}{J\alpha {{\beta }_{h}}{{\beta }_{g}}}}} )>0$. Therefore, $R_s$ increases monotonically with ${{\left| \mathbf{w}_{d}^{\text{H}}{{\mathbf{h}}_{d}} \right|}^{2}}$.
In other words, maximizing $R_s$ is equivalent to maximizing ${{\left| \mathbf{w}_{d}^{\text{H}}{{\mathbf{h}}_{d}} \right|}^{2}}$, which is obviously given by $\mathbf{w}_{d}=\frac{{{\mathbf{h}}_{d}}}{\left\| {{\mathbf{h}}_{d}} \right\|}$.
\end{IEEEproof}
Based on Lemma 1 and Lemma 3, by eliminating the common variable $J$, the asymptotic primary communication rate $R_s$ can be expressed in closed-form in terms of the secondary sum-rate ${R}_{BD}$ as:
\begin{equation}
\setlength\abovedisplayskip{0.2cm}
\setlength\belowdisplayskip{0.2cm}
{{{R}_{s}}={{\log }_{2}}\gamma -\text{Ei}\left( -\frac{K\gamma }{{{2}^{\frac{K}{M}{{R}_{BD}}}}-1} \right){{\log }_{2}}e,}\label{eq:RSRBD}
\end{equation}
where $\gamma \triangleq \frac{p{{\left\| {{\mathbf{h}}_{d}} \right\|}^{2}}}{{{\sigma }^{2}}}$ is the maximum direct-link SNR. It is not difficult to show that $R_s$ in \eqref{eq:RSRBD} monotonically increases with ${R}_{BD}$, which clearly reveals the mutualism relationship of symbiotic radio with massive BDs.
\begin{figure}[!t]
  \setlength{\abovecaptionskip}{-0.1cm}
  \setlength{\belowcaptionskip}{-0.2cm}
  \centering
  \centerline{\includegraphics[width=3.0in,height=2.5in]{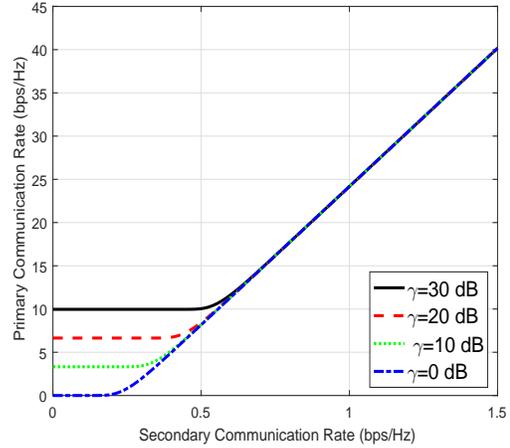}}
  \caption{The asymptotic performance of primary versus secondary communication rate.}
  \label{system model}
   \vspace{-0.5cm}
  \end{figure}

Fig. 2 gives an example plot of \eqref{eq:RSRBD} with $M=4$ and $K=128$, for four different values of $\gamma$. It is worth mentioning that while \eqref{eq:RSRBD} was derived for asymptotic setup with $J\gg 1$, it is also applicable for the extreme case with $J=0$ or ${R}_{BD}=0$, for which case the second term in \eqref{eq:RSRBD} vanishes. Therefore, Fig. 2 plots $R_s$ versus ${R}_{BD}$ in \eqref{eq:RSRBD}, starting from ${R}_{BD}=0$. It is observed from Fig.  2 that when the secondary communication rate is small, increasing ${R}_{BD}$ (or equivalently increasing $J$) has negligible impact on the primary communication rate. However, once ${R}_{BD}$ or $J$ exceeds certain threshold, $R_s$ increases almost linearly with ${R}_{BD}$. This thus demonstrates that the mutualism of symbiotic radio can only be fully exploited for sufficient BDs. It is also interesting to observe that as ${R}_{BD}$ becomes sufficiently large, $R_s$ for different $\gamma$ values merge. This can be verified by taking the derivative of $R_s$ in \eqref{eq:RSRBD} with respect to $\gamma$, which vanishes as ${R}_{BD}$ gets sufficiently large.
%Take $\gamma$ as a variable to derive $R_s$ in (27), we have $\frac{\partial {{R}_{s}}}{\partial \gamma }={{\log }_{2}}e\left( \frac{1-{{e}^{\left( \frac{-K\gamma }{{{2}^{\frac{K}{M}{{R}_{BD}}}}-1} \right)}}}{\gamma } \right)$. As ${R}_{BD}$ increases, $\frac{-K\gamma }{{{2}^{\frac{K}{M}{{R}_{BD}}}}-1}\to 0$, $\frac{\partial {{R}_{s}}}{\partial \gamma }\to 0$, which means $R_s$ is independent of $\gamma$. This explains why the four curves in Fig. 2 merge as ${R}_{BD}$ increases.

Next, we consider the beamforming optimization problem for the generic case for small or moderate number of BDs $J$. With the semi-closed form expression (20), (P1) can be reformulated as
\setlength{\skip\footins}{2ex}
\begin{small}
\begin{align}
\setlength\abovedisplayskip{2pt}
\setlength\belowdisplayskip{2pt}
\notag\text{(P3)}\  \underset{{{\left\| {{\mathbf{w}}_{d}} \right\|}}=1}{\mathop{\max }}\,& \text{ }\text{lo}{{\text{g}}_{2}}\frac{p{{\left| \mathbf{w}_{d}^{\text{H}}{{\mathbf{h}}_{d}} \right|}^{2}}}{{{\sigma }^{2}}}-\text{Ei}\Big( -\frac{p{{\left| \mathbf{w}_{d}^{\text{H}}{{\mathbf{h}}_{d}} \right|}^{2}}}{\sum\limits_{j=1}^{J}{p\alpha {{\left| {{h}_{j}} \right|}^{2}}{{\left| \mathbf{w}_{d}^{\text{H}}{{\mathbf{g}}_{j}} \right|}^{2}}}} \Big){{\log }_{2}}e.
\end{align}
\end{small}Let ${{\mathbf{H}}_{d}}=p{{\mathbf{h}}_{d}}\mathbf{h}_{d}^{\text{H}}$, ${{\mathbf{H}}_{j}}=p\alpha {{\left| {h_j} \right|}^{2}}{{\mathbf{g}}_{j}}\mathbf{g}_{j}^{\text{H}}$, and ${{\mathbf{W}}_{d}}={{\mathbf{w}}_{d}}\mathbf{w}_{d}^{\text{H}}$. (P3) can be recast as
\begin{small}
\begin{align}\label{Optimization Problem3}
\setlength\abovedisplayskip{2pt}
\setlength\belowdisplayskip{2pt}
\notag\text{(P3-SDP)}\     \underset{\text{}}{\mathop{\underset{{{\mathbf{W}}_{d}}}{\mathop{\max }}\,}}\,& \text{    }\text{lo}{{\text{g}}_{2}}\frac{\text{Tr}\left( {{\mathbf{W}}_{d}}{{\mathbf{H}}_{d}} \right)}{{{\sigma }^{2}}}-\text{Ei}\left( -\frac{\text{Tr}\left( {{\mathbf{W}}_{d}}{{\mathbf{H}}_{d}} \right)}{\text{ }\sum\limits_{j=1}^{J}{\text{Tr}\left( {{\mathbf{W}}_{d}}{{\mathbf{H}}_{j}} \right)}} \right){{\log }_{2}}e \\
  \text{ s}\text{.t.    }& {{\mathbf{W}}_{d}}\succeq0,\text{Tr}\left( {{\mathbf{W}}_{d}} \right)=1,\\
  & \text{Rank}\left( {{\mathbf{W}}_{d}} \right)=1.
\end{align}
\end{small}To solve (P3-SDP), we follow the similar technique as~\cite{IEEEhowto:7}, by introducing an auxiliary variable $\xi \triangleq \frac{\text{Tr}\left( {{\mathbf{W}}_{d}}{{\mathbf{H}}_{d}} \right)}{\text{ }\sum\limits_{j=1}^{J}{\text{Tr}\left( {{\mathbf{W}}_{d}}{{\mathbf{H}}_{j}} \right)}}$.
By relaxing the nonconvex rank-one constraint (25), we recast problem (P3-SDP) as
\begin{small}
\begin{align}\label{Optimization Problem3}
\setlength\abovedisplayskip{2pt}
\setlength\belowdisplayskip{2pt}
   \underset{\text{                         }}{\mathop{\text{(P3-SDR)     }\underset{{{\mathbf{W}}_{d}},\xi }{\mathop{\max }}\,}}\,& \text{    }\text{lo}{{\text{g}}_{2}}\frac{\text{Tr}\left( {{\mathbf{W}}_{d}}{{\mathbf{H}}_{d}} \right)}{{{\sigma }^{2}}}-\text{Ei}\left( -\xi  \right){{\log }_{2}}e \\
  \text{                     s}\text{.t.   }& {{\mathbf{W}}_{d}}\succeq0,\text{Tr}\left( {{\mathbf{W}}_{d}} \right)=1,\\
  & \text{Tr}\left( {{\mathbf{W}}_{d}}{{\mathbf{H}}_{d}} \right)-\xi \sum\limits_{j=1}^{J}{\text{Tr}\left( {{\mathbf{W}}_{d}}{{\mathbf{H}}_{j}} \right)}=0.
\end{align}
\end{small}For any fixed $\xi$, (P3-SDR) is a semidefinite optimization problem, which can be optimally solved by using software tools like CVX \cite{IEEEhowto:22}. Then the optimal ${\xi}^{\star}$ can be obtained by one-dimensional exhaustive search over $\xi$. After obtaining the solution $\mathbf{W} ^{\star}$ to (P3-SDR), we can use the standard Gaussian randomization procedures \cite{IEEEhowto:23} to find the receive beamforming solution $\mathbf{w}_{d} ^{\star}$ to (P3).
\section{Simulation Results}
In this section, simulation results are provided to evaluate the performance of the studied symbiotic radio system. We assume independent random channels for all communication links, where the small-scale fading components follow i.i.d. CSCG distribution with zero mean and unit-variance, and the large-scale channel gains for the PT-to-AP, PT-to-BD, and BD-to-AP channels are $\beta_{hd}=-120$ dB, $\beta_h=-110$ dB, and $\beta_g=-20$ dB, respectively, i.e., ${\mathbf {h}_d}\sim \mathcal{C}\mathcal{N}(\mathbf 0, \beta_{hd}\mathbf I_M)$, ${{h}_{j}}\sim \mathcal{C}\mathcal{N}(0,\beta_h)$, and $\mathbf g_j\sim \mathcal{CN}(\mathbf 0, \beta_g \mathbf I_M)$, for $j=1,2,\cdots ,J$. The noise power is $\sigma^2=-110$ dBm, the number of receive antennas at the AP is $M=4$, and the power reflection coefficient is $\alpha =1$. Furthermore, we set the ratio between the symbol duration of the BD symbols and that of the PT symbols as $K=128$.

\begin{figure}[!t]
  \setlength{\abovecaptionskip}{-0.1cm}
  \setlength{\belowcaptionskip}{-0.2cm}
  \centering
  \centerline{\includegraphics[width=3.1in,height=2.5in]{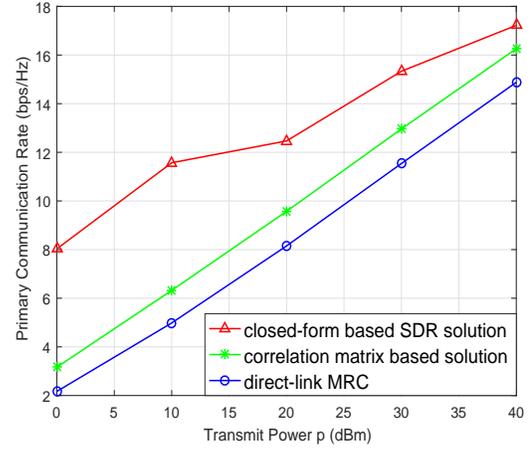}}
  \caption{Average primary communication rate versus transmit power.}
  \label{system model}
   \vspace{-0.3cm}
  \end{figure}
  \begin{figure}[!t]
  \setlength{\abovecaptionskip}{-0.1cm}
  \setlength{\belowcaptionskip}{-0.2cm}
  \centering
  \centerline{\includegraphics[width=3.1in,height=2.5in]{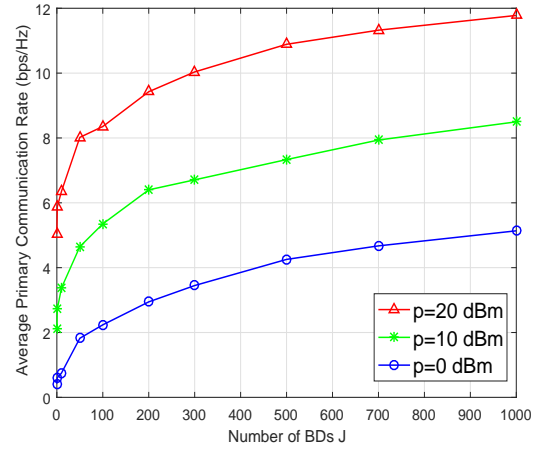}}
  \caption{Average primary communication rate versus number of BDs $J$.}
  \label{system model}
   \vspace{-0.3cm}
  \end{figure}
\begin{figure}[!t]
  \setlength{\abovecaptionskip}{-0.1cm}
  \setlength{\belowcaptionskip}{-0.2cm}
  \centering
  \centerline{\includegraphics[width=3.1in,height=2.5in]{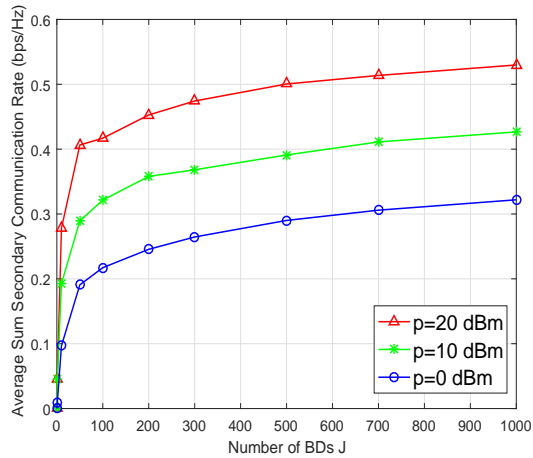}}
  \caption{Average secondary communication rate versus number of BDs $J$.}
  \label{system model}
   \vspace{-0.3cm}
  \end{figure}
Fig. 3 compares the performance of the primary communication rate with three different beamforming schemes. Besides the correlation matrix based solution proposed in subsection IV A and the closed-form expression based SDR solution proposed in subsection IV B, we also consider the direct-link maximal-ratio combining (MRC) beamforming scheme as a benchmark, where the AP ignores the multipath created by the BDs and simply sets the receive beamforming to match the direct link, i.e., $\mathbf{w}_{d}=\frac{{{\mathbf{h}}_{d}}}{\left\| {{\mathbf{h}}_{d}} \right\|}$. The results in Fig. 3 are obtained for one realization of the channels, and the number of BDs is $J=200$. It is observed from Fig. 3 that for all the three beamforming schemes, the primary rate increases monotonically with the transmit power $p$, as expected. Furthermore, both our proposed solutions in Section IV outperform the direct-link MRC beamforming, thanks to the consideration of the effective channel constituted by both the direct link and the backscattered multipaths. Furthermore, it is observed that the closed-form expression based SDR solution gives the best performance, but at the cost of higher computation complexity since it requires solving a sequence of SDP problems and one-dimensional exhaustive search.

Next, we study the impact of the number of BDs $J$ on the average primary and secondary communication rates, where the average is taken over $1000$ independent channel realizations. For each channel realization, the sum of the secondary communication rate is obtained based on the closed-form expression (13), and the primary communication rate is obtained with the correlation matrix based solution. Note that though sub-optimal in general, the correlation matrix based solution has much lower complexity than the closed-form expression based SDR solution, and it provides useful performance lower bound. Fig. 4 and Fig. 5 plot the average primary and secondary communication rates versus the number of BDs $J$, respectively. It is firstly observed that the primary communication rate is in general much higher than the secondary rate. This is expected since the symbol rate of primary signals is $K=128$ times of that of the secondary signals, and that the backscattered link for one single BD is in general much weaker than the primary communication link. Furthermore, it is observed that as $J$ increases, both primary and secondary rates increase, which corroborate our theoretical study in Section III. As a concrete example, consider the primary rate in Fig. 4. Compared to the case without any BD $(J=0)$, the rate improvement for one single BD $(J=1)$ when $p=0$ dBm is only 0.19 bps/Hz, while that for $J=500$ is 3.83 bps/Hz. Such results demonstrate that the full mutualism relationship of symbiotic radio can be enabled by massive number of BDs.
\section{Conclusion}\label{conclusion}
In this paper, we studied symbiotic radio systems with multiple BDs to fully exploit the mutualism between primary and secondary transmissions. We first derive the achievable rate of both the primary and secondary communications, based on which a
receive beamforming optimization problem is formulated and solved. Furthermore, considering the asymptotic regime of massive number of BDs, closed-form expressions are derived for the primary and secondary communication rates, both of which are shown to be increasing functions of the number of BDs. This thus demonstrates that the mutualism relationship of symbiotic radio can be enhanced with massive BD access. Simulation results were provided to validate our theoretical studies.
\section{Acknowledgment}
This work was supported by the National Key R\&D Program of China with grant number 2019YFB1803400.
\end{document}